\documentclass[twocolumn,trackchanges]{aastex701}
\usepackage{xspace}
\usepackage{amsmath}
\usepackage{multirow} 
\newcommand{\hxmt}{\textit{Insight}-HXMT\xspace}
\newcommand{\us}{$\mu$s\xspace}
\newcommand{\um}{$\mu$m\xspace}

\usepackage{bm,xcolor,color}

\definecolor{mycolor}{rgb}{0,0,0}
\newcommand{\revt}[1]{#1}

\newcommand{\revtt}[1]{#1}

\usepackage{CJK}

\begin{document}
\begin{CJK*}{UTF8}{gbsn}

\title{In-flight Timing Calibration of the Follow-up X-ray Telescope onboard Einstein Probe}

\correspondingauthor{Juan Zhang}
\email[]{zhangjuan@ihep.ac.cn}

\correspondingauthor{Hao Wang}
\email[]{wanghao87@ihep.ac.cn}

\author[orcid=0000-0001-8869-0672,gname=Juan, sname=Zhang]{Juan Zhang(张娟)}
\affiliation{State Key Laboratory of Particle Astrophysics, Institute of High Energy Physics, Chinese Academy of Sciences, Beijing, 100049, China}
\email[]{zhangjuan@ihep.ac.cn}  

\author[orcid=,gname=Xiaofan, sname=Zhao]{Xiaofan Zhao(赵晓帆)}
\affiliation{State Key Laboratory of Particle Astrophysics, Institute of High Energy Physics, Chinese Academy of Sciences, Beijing, 100049, China}
\email[]{zhaoxf@ihep.ac.cn}

\author[orcid=,gname=Dawei, sname=Han]{Dawei Han(韩大炜)}
\affiliation{State Key Laboratory of Particle Astrophysics, Institute of High Energy Physics, Chinese Academy of Sciences, Beijing, 100049, China}
\email[]{dwhan@ihep.ac.cn}

\author[orcid=,gname=Hao, sname=Wang]{Hao Wang(王昊)}
\affiliation{State Key Laboratory of Particle Astrophysics, Institute of High Energy Physics, Chinese Academy of Sciences, Beijing, 100049, China}
\email[]{wanghao87@ihep.ac.cn}

\author[orcid=,gname=Yong, sname=Chen]{Yong Chen(陈勇)}
\affiliation{State Key Laboratory of Particle Astrophysics, Institute of High Energy Physics, Chinese Academy of Sciences, Beijing, 100049, China}
\email[]{ychen@ihep.ac.cn} 

\author[orcid=,gname=Mingyu, sname=Ge]{Mingyu Ge(葛明玉)}
\affiliation{State Key Laboratory of Particle Astrophysics, Institute of High Energy Physics, Chinese Academy of Sciences, Beijing, 100049, China}
\email[]{gemy@ihep.ac.cn}  

\author[orcid=,gname=Shijie, sname=Zheng]{Shijie Zheng(郑世界)}
\affiliation{State Key Laboratory of Particle Astrophysics, Institute of High Energy Physics, Chinese Academy of Sciences, Beijing, 100049, China}
\email[]{zhengsj@ihep.ac.cn}

\author[orcid=,gname=Shumei, sname=Jia]{Shumei Jia(贾淑梅)}
\affiliation{State Key Laboratory of Particle Astrophysics, Institute of High Energy Physics, Chinese Academy of Sciences, Beijing, 100049, China}
\email[]{jiasm@ihep.ac.cn}

\author[orcid=,gname=Haisheng, sname=Zhao]{Haisheng Zhao(赵海升)}
\affiliation{State Key Laboratory of Particle Astrophysics, Institute of High Energy Physics, Chinese Academy of Sciences, Beijing, 100049, China}
\email[]{zhaohs@ihep.ac.cn} 

\author[orcid=,gname=Weiwei, sname=Cui]{Weiwei Cui(崔苇苇)}
\affiliation{State Key Laboratory of Particle Astrophysics, Institute of High Energy Physics, Chinese Academy of Sciences, Beijing, 100049, China}
\email[]{cuiww@ihep.ac.cn}

\author[orcid=,gname=Hanlong, sname=Peng]{Hanlong Peng(彭寒龙)}
\affiliation{School of Physics and Technology, Nanjing Normal University, Nanjing, 210023, Jiangsu, China}
\email[]{penghl@ihep.ac.cn} 

\author[orcid=,gname=Wentao, sname=Ye]{Wentao Ye(叶文韬)}
\affiliation{State Key Laboratory of Particle Astrophysics, Institute of High Energy Physics, Chinese Academy of Sciences, Beijing, 100049, China}
\email[]{yewt@mail.ihep.ac.cn} 

\author[orcid=,gname=Yaoming, sname=Duan]{Yaoming Duan(段耀明)}
\affiliation{State Key Laboratory of Particle Astrophysics, Institute of High Energy Physics, Chinese Academy of Sciences, Beijing, 100049, China}
\email[]{ymduan@ihep.ac.cn} 

\author[orcid=,gname=Youli, sname=Tuo]{Youli Tuo(庹攸隶)}
\affiliation{Institut f\"ur Astronomie und Astrophysik, Universit\"at T\"ubingen, Sand 1, D-72076 T\"ubingen, Germany}
\email[]{youli.tuo@astro.uni-tuebingen.de}

\author[orcid=,gname=Zhen, sname=Ji]{Zhen Ji(纪珍)}
\affiliation{National Space Science Center, Chinese Academy of Sciences, Beijing 100190, China}
\email[]{zji@nssc.ac.cn} 

\author[orcid=,gname=Fuli, sname=Ma]{Fuli Ma(马福利)}
\affiliation{National Space Science Center, Chinese Academy of Sciences, Beijing 100190, China}
\email[]{mafuli007@nssc.ac.cn} 

\author[orcid=,gname=Chengkui, sname=Li]{Chengkui Li(李承奎)}
\affiliation{State Key Laboratory of Particle Astrophysics, Institute of High Energy Physics, Chinese Academy of Sciences, Beijing, 100049, China}
\email[]{lick@ihep.ac.cn}

\author[orcid=,gname=Ju, sname=Guan]{Ju Guan(关菊)}
\affiliation{State Key Laboratory of Particle Astrophysics, Institute of High Energy Physics, Chinese Academy of Sciences, Beijing, 100049, China}
\email[]{jguan@ihep.ac.cn}

\author[orcid=,gname=Jingjing, sname=Xu]{Jingjing Xu(徐静静)}
\affiliation{State Key Laboratory of Particle Astrophysics, Institute of High Energy Physics, Chinese Academy of Sciences, Beijing, 100049, China}
\email[]{xujingjing@ihep.ac.cn}

\author[orcid=,gname=Jin, sname=Wang]{Jin Wang(王进)}
\affiliation{State Key Laboratory of Particle Astrophysics, Institute of High Energy Physics, Chinese Academy of Sciences, Beijing, 100049, China}
\email[]{jinwang@ihep.ac.cn}

\author[orcid=,gname=Liming, sname=Song]{Liming Song(宋黎明)}
\affiliation{State Key Laboratory of Particle Astrophysics, Institute of High Energy Physics, Chinese Academy of Sciences, Beijing, 100049, China}
\email[]{songlm@ihep.ac.cn}




\begin{abstract}

After the launch of Einstein Probe (EP), pulsars are observed to calibrate the timing system of the onboard Follow-up X-ray Telescope (FXT) by measuring the times of arrival (TOAs). FXT has two independently operating modules, namely FXTA and FXTB, each equipped with the Wolter-I optics and the pnCCD camera. For scientific observations, each pnCCD has three distinct readout modes that differ in frame cycles, the full frame mode (FF, 50\,ms per frame), the partial windowed mode (PW, 2.2\,ms per frame), and the timing mode (TM, 23.68\,$\mu$s per row, 9.4352\,ms per frame). Thus, 1RXS J170849.0--400910 with the spin period of 11\,s is observed in FF, PSR B1509--58 (150\,ms) in PW, and Crab pulsar (33\,ms) in TM as well as PSR B1821--24 (3.05\,ms) and PSR B1937+21 (1.56\,ms). The TOAs of these pulsars measured by FXTA and FXTB are compared with the results of NICER, and for Crab pulsar with the Jodrell Bank Monthly Ephemeris (JBE). Results verify that the timing system of FXTA and FXTB in FF, PW, and TM modes are consistent with each other, and are in good alignment with NICER. The measured advance of the Crab pulsar's main peak relative to JBE is 
\revt{
$357^{\pm14(stat)}_{\pm54(sys)}$\,\us in FXTA and $394^{\pm14(stat)}_{\pm67(sys)}$\,\us in FXTB. Based on measurements of PSR B1821--24 and PSR B1937+21, the preliminary absolute timing accuracy is estimated to be 6\,\us for FXTA and 31\,\us for FXTB in TM mode. Additionally, the systematic delay of FXTA timing results relative to FXTB is found to be $20^{\pm19(stat)}_{\pm21(sys)}$\,\us.
}

\end{abstract}


\section{Introduction}\label{sec:intro} 

Einstein Probe (EP), launched into a low earth orbit (LEO) on 2024 January 9, is a mission spearheaded by the Chinese Academy of Sciences (CAS) in collaboration with the European Space Agency (ESA), the Max Planck Institute for Extraterrestrial Physics (MPE), and the Centre National d'Etudes Spatiales (CNES) \citep{2022hxga.book...86Y}. Its primary scientific objective is to investigate time-domain X-ray astrophysics by studying cosmic high-energy transients and monitoring variable celestial objects. This is accomplished by carrying two types of payloads, the Wide-field X-ray Telescope (WXT, 0.5--4\,keV, \citealt{2024ExA....57...10C}) and the Follow-up X-ray Telescope (FXT, 0.3--10\,keV, \citealt{2025RDTM....9..198C}). WXT employs the lobster-eye micro-pore optics and CMOS detectors. And FXT utilizes the Wolter-I optics plus pnCCD cameras with a focal length of 1.6\,m and spatial resolution of 30$^{\prime\prime}$. With the wide field-of-view (FOV) of 3600 square degrees ($\sim$1.1\,Sr) and high sensitivity \citep{2017ExA....43..267Z}, WXT has the advantage to capture transients and to monitor variable celestial objects extensively. After trigger, FXT starts the follow-up observations and astrophysical investigations by virtue of its larger effective area, better spatial and temporal resolutions \citep{2023ExA....55..603C,2023ExA....55..625Y,2025RDTM....9..198C,2025RAA....25a5002Z}. 

The precise timing capabilities of instruments play a crucial role in advancing our understanding of the dynamic behavior and physical processes of a wide range of astrophysical sources, such as the millisecond pulsars (MSPs, e.g. \citealt{2013ApJ...762...96B,2019ApJ...878L..22R,2025ApJ...981...99Q,2025ApJ...990...15L}), fast radio bursts, magnetars and quasi-periodic oscillations \citep{2021NatAs...5..378L,2022ApJ...931...56L}, X-ray binaries \citep{2025ApJ...991L..41C}, and tidal disruption events \citep{2021ApJ...920...60J,2022MNRAS.511L..19Z}.
Before launch, the on-ground timing calibration of EP/FXT in TM mode \citep{2025RAA....25a5002Z} has verified that: the relative timing accuracy exceeds $5\times10^{-9}$; the time resolution is about $46$\,\us; and the absolute time delay is $\sim4\pm2$\,\us. This work intends to investigate the timing performance of EP/FXT after launch. 

Conventionally, the in-flight timing calibration for X-ray and gamma-ray instruments was performed through comparative analysis of times of arrival (TOAs) of pulsars against reference timing ephemerides. The timing accuracy of both XMM-Newton/EPIC-pn and Swift/XRT were obtained by analysing the observations of Crab pulsar \citep{2012A&A...545A.126M,2012A&A...548A..28C} in comparison with the Jodrell Bank Crab Pulsar Monthly Ephemeris\footnote{\url{http://www.jb.man.ac.uk/~pulsar/crab.html}} (JBE, \citealt{1993MNRAS.265.1003L}). Considering the limited systematic errors of JBE, \citet{2022ApJS..259...14T} analysed the quasi-simultaneous observations of the Neutron Star Interior Composition Explorer (NICER) to obtain the reference ephemeris of Crab pulsar and to verify the timing system of \hxmt. In the International Astronomical Consortium for High Energy Calibration (IACHEC) timing cross-calibration paper \citep{2026PASP..138a4501B}, a new method was developed to compare the pulsar TOAs observed among different missions and epochs. It is based on a reference timing solution, e.g. JBE for Crab pulsar, simulates a series of TOAs, and then fits the new timing solutions as reference to be used in cross-calibration. 

In this work, we perform the in-flight timing calibration for the three readout modes of each EP/FXT module, by using pulsars with spin periods spanning different time scales as references. This paper is organized as follows. In Section\,\ref{sec:fxttime}, we describe the time system of EP/FXT and time tagging of events in each mode. The local clock stability after launch is depicted in Section\,\ref{sec:localclock}. And pulsar observations and data analysis are introduced in Section\,\ref{sec:obs}. The timing calibration results are presented in Section\,\ref{sec:results}, and summarized in Section\,\ref{sec:con}. 
 
\section{Time System and Event Time Tagging}\label{sec:fxttime}

EP carries a Global Navigation Satellite System (GNSS) receiver, which obtains accurate time signal from  Global Positioning System and/or Beidou\footnote{The Chinese satellite navigation system} and distributes the corresponding Pulse per Second (PPS) to FXT modules. PPS is followed by a Coordinated Universal Time (UTC) event. The PPS and UTC events are stored together as a pair in FXT raw data package. FXTA and FXTB utilize individual 100\,MHz temperature-compensated crystal oscillators to acquire their respective local clock with a nominal cycle of 0.1\,\us \citep{2025RAA....25a5002Z}. The readout electronics in each module uses its own crystal oscillator to count the local time independently. 

The pnCCDs of FXTA and FXTB modules are arranged orthogonally on the focal plane and are configured with three readout modes for observing astrophysical sources with diverse luminosities and timing properties. These three modes are referred to as the full-frame mode (FF), the partial-window mode (PW), and the timing mode (TM) \citep{2023ExA....55..603C}. 
\begin{itemize}
    \item In FF, all 384$\times$384 pixels of the pnCCD image area (IMG) are readout as a frame per 50\,ms, of which 49.88485\,ms for exposure and 0.11515\,ms for shift from IMG to the frame store area (FSA) with a fast transfer speed of 0.3\,\us per row. Prior to reading out the first row, a wait time of 4.28485\,ms occurs \citep{2025RAA....25k5019Z}. Consequently, a time offset of 4.4\,ms arises between the start of the frame shift from IMG to FSA and the readout of the first row when the frame is time-stamped with the local time count.     
    \item Differently, PW only reads the central 128 columns $\times$ 61 rows per 2.2\,ms as a frame, which also uses fast transfer from IMG to FSA. The time difference between the start of the frame shift and the begin of readout is 517\,\us. 
    \item In TM, the central 128 columns$\times$384 rows are slowly shifted and readout continuously at a rate of 23.68\,$\mu$s per row; after readout of these 384 rows of a frame, an additional recover time of 342.08\,$\mu$s is required to reset the electronics; thus the frame cycle of TM is 9.4352\,ms, which is also the time difference between the start of the shift and its readout. 
\end{itemize}

The local time count is recorded in each frame of FF, PW and TM. The PPS and UTC events are inserted into FXT data stream, with PPS tagged with local time count as well. The time of each frame is decoded following the similar method as \hxmt \citep{2025RAA....25a5002Z,2022ApJS..259...14T}. The PPS time and their corresponding local time count are combined to give a best fit so as to compensate the potential drift of the local crystal oscillator clock and the potential errors of PPS, 
\begin{equation}
\label{eq:fit_local_clock}
t = t{_0} + \tau(N-N{_0}) + \frac{1}{2}\dot{\tau}(N-N{_0}){^2}  + \frac{1}{6}\ddot{\tau}(N-N{_0}){^3} + ...,
\end{equation}
where t is the time of PPS event, N is the corresponding local time count, $N{_0}$ and $t{_0}$ are those of the first PPS event in this interval, $\tau$ is the actual period of the local time cycle, and $\dot{\tau}$ and $\ddot{\tau}$ the first- and second-order derivatives. Given the time resolution of FXT, a linear model is sufficiently accurate for in-orbit data fitting over a 50-second interval (ground calibration used a 100-second interval, \citealt{2025RAA....25a5002Z}).

After each time interval fitting, the obtained parameters are used to determine the time of each frame with their known local time count. This frame time is subtracted by the aforementioned time difference between the start of shift of each frame and its readout, yielding the actual frame time that is written in the Level-1 (L1, unfiltered) data of FXT. Events within the same frame share the frame time. In FF and PW modes, a fast shift is used with the exposure time \(\gg\) shift time, so the arrival times of events cannot be distinguished within a single frame and are assigned a common time tag at the frame level. By contrast, TM mode uses a slow shift with continuous readout and a reset time that is very short. Furthermore, relative to the pnCCD size (IMG 384$\times$384 pixels, each pixel 75\,$\mu$m$\times$75\,\um), the FXT spatial resolution, $\sim30^{\prime\prime}$ which corresponds to about 3 pixels in diameter, is much smaller. The source photons mainly strike only the focal-spot region during shifting and resetting. As the signals in rows are shifted, the time information of celestial photons incident at different instants is transformed to the spatial information recorded at different rows on the pnCCD. Consequently, events in TM undergo time corrections as depicted in Equation\,(\ref{eq:tm}) during the reduction process from L1 to Level-2 (L2, cleaned) data \citep{2025RDTM....9..215Z}, yielding improved time resolution within the frame period, 
\begin{widetext}
\begin{equation}
    t_{evt}(row) = 
\begin{cases} 
t_{frame} + 236.8\tau(row -row0), & \text{when $row$ $\ge$ $row0$;} \\
t_{frame} + 236.8\tau(row -row0) - 3420.8\tau, & \text{when $row$ $<$ $row0$;}
\end{cases}
\label{eq:tm}
\end{equation}
\end{widetext}
where $t_{frame}$ is the actual frame time in L1, $row$ the row indices recorded by events, $row0$ the source row position in the detector coordinate, and $\tau$ the local time cycle at that time. Roughly speaking, the photon time was corrected by the difference in row indices between the target source position and the recorded events on the detector coordinate. For events located below the source position, i.e. with $row<row0$, they hit on the detector during the transfer of previous frame from IMG to FSA, thus an additional reset time between frames is additionally subtracted. The ground timing calibration \citep{2025RAA....25a5002Z} has verified the validity of this reduction method for event times of TM mode.

\section{Local Clock Stability\label{sec:localclock}}

\begin{figure*}[htb]
    \centering
    \includegraphics[width=0.45\linewidth]{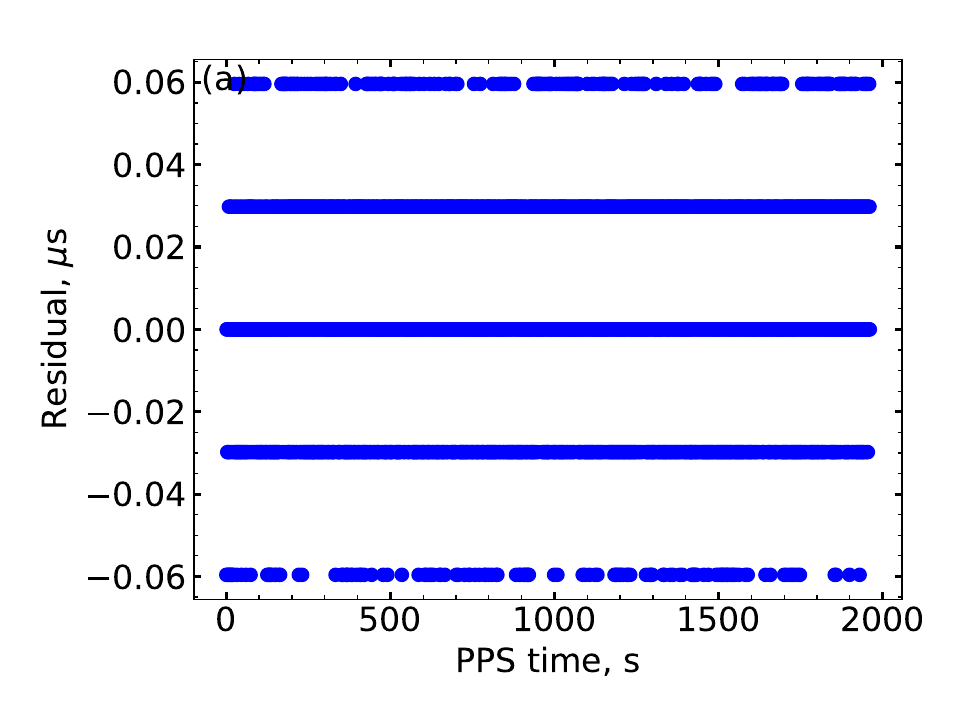}
    \includegraphics[width=0.45\linewidth]{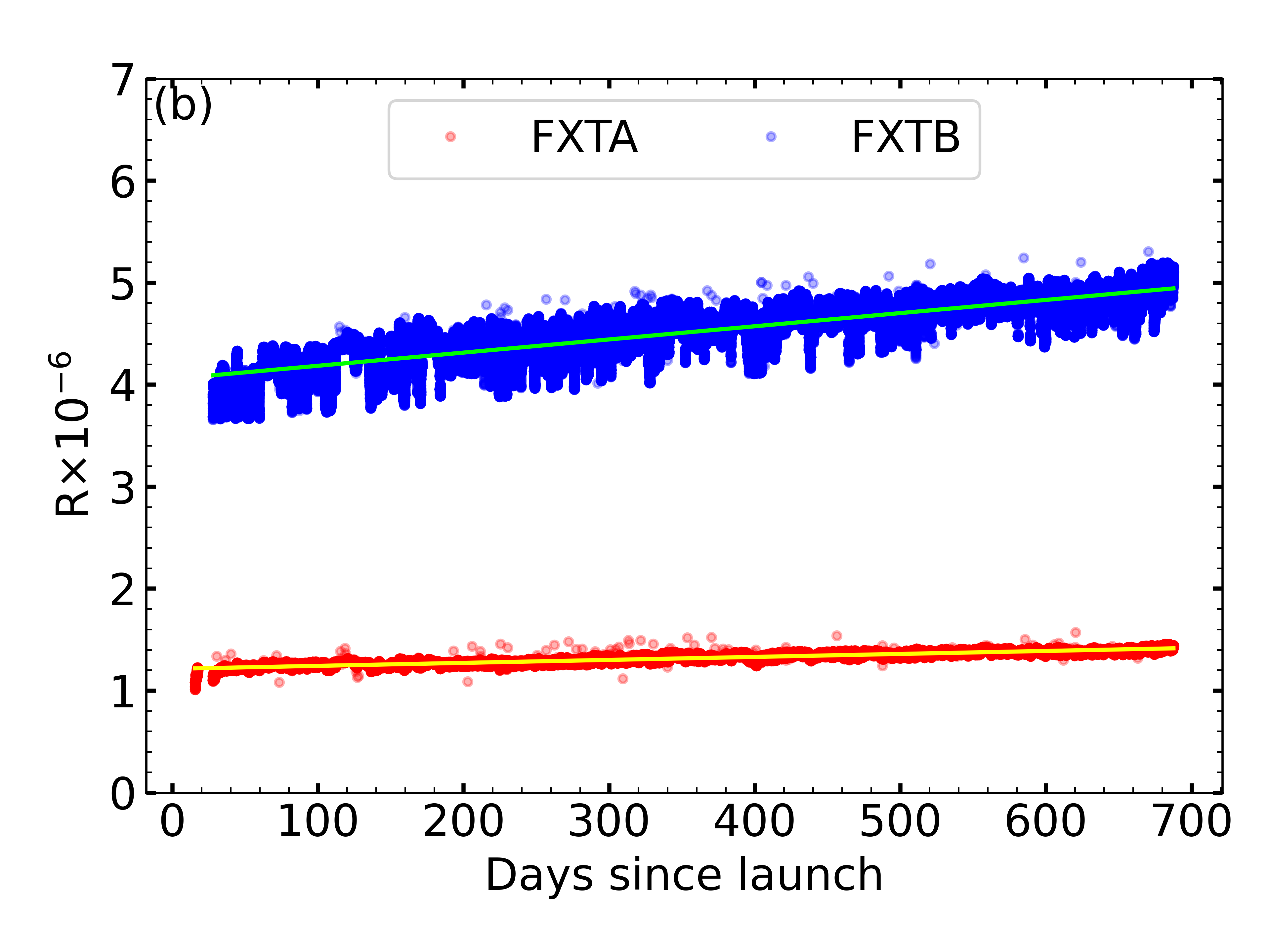}
    \caption{Panel\,(a): Residuals between the decoded PPS event times and the PPS reference times, using a single FXTA observation as an example. Panel\,(b): The relative evolution of the local time cycle of the crystal oscillator in FXTA (red) and FXTB (blue), with the corresponding linear trends modeled for FXTA (yellow) and FXTB (lime).\label{fig:gpsresidual_and_cyclechange}}
\end{figure*}

The residuals between the decoded time of PPS events and the PPS time are illustrated in  Figure\,\ref{fig:gpsresidual_and_cyclechange}(a). The differences are less than the nominal cycle of the local crystal oscillator clock, i.e. 0.1\,\us. 
Figure\,\ref{fig:gpsresidual_and_cyclechange}(b) shows the relative evolution of the local time cycle of FXTA and FXTB, respectively, since launch to the end of 2025 November. The relative change of the local clock cycle is defined the same as \hxmt (Equation\,(3) in \citealt{2022ApJS..259...14T}), which is  
\begin{equation}
    R = \frac{\tau - \tau_0}{\tau_0},
\end{equation}
where $\tau$ is the actual period of the local cycle, and $\tau_0$ the nominal local cycle which is  0.1\,\us for FXT. This plot clearly shows that the relative deviation from the nominal cycle is on the order of $10^{-6}$ for both FXTA and FXTB. The crystal oscillators of FXTA and FXTB behave differently:
\begin{itemize}
\item the local clock cycle of FXTB is longer than that of FXTA;
\item the relative change in FXTA follows a secular drift of  $\sim2.9\times10^{-10}$ per day, and FXTB $\sim1.3\times10^{-9}$ per day;
\item the RMS in FXTA is 0.07$\times10^{-6}$ after taking the secular drift into account, and in FXTB 0.15$\times10^{-6}$.
\end{itemize}
It cloud be concluded that the local clock of FXTA is more stable than FXTB after launch.

\section{Observations and Data Analysis} \label{sec:obs}

\subsection{\revt{FXT Observations and Data Reduction}}

\begin{table*}[htb]   
\caption{The in-flight observation logs of FXT timing calibration till 2025 mid\label{tab:obs} }
\begin{tabular}{c|ccccc}
\hline
\hline
\textbf{Mode} & \textbf{Source} &  \textbf{obsID} & \textbf{start date (UTC)} & \textbf{start MJD} & \textbf{exposure, ks} \\ 
              &                 &                 &      YYYY-MM-DD           & {}  &    FXTA/FXTB          \\
\hline
\textbf{FF}       &  \textbf{RXJ1708}   &   13600005153     &  2024-04-06                 &    60406 &    33.3/7.9      \\
\textbf{} &    \textbf{ }              &   05100000001     &  2024-05-11        &    60441         &         86.7         \\
\hline
\textbf{PW} &  \textbf{PSR B1509--58} & 13600005956    &  2024-05-07    &    60437             &         47.9          \\
\hline
        &        &   08500000012       &  2024-02-28  &    60368    &     --/2.6   \\
        &    \textbf{}     &   13600005102     &  2024-03-08      &    60377 & 8.3\\
        &                         &   11900008151     &  2024-09-23   &    60576    & 5.9\\
        &  \textbf{Crab}                       &   11900008152     &  2024-09-23   &    60576   & 5.9\\            
\textbf{TM}       &                         &   08500000313  &  2025-03-22  & 60756        & 2.0\\
\textbf{}       &                         &   08500000314     &  2025-03-22 & 60756       & 2.9\\
        &                         &   08500000324     &  2025-03-24   & 60758    & 3.0\\
        \cline{2-6}
        &  \textbf{PSR B1821--24}  &   13600005957     &  2024-05-08    & 60438  &  54.2   \\
        &  \textbf{}    &   13600005958     &  2024-05-10   & 60440    &  55.3   \\    
        \cline{2-6}
        &  \textbf{PSR B1937+21}  &   05100000003     &  2024-05-13  & 60443    &  66.3   \\
\hline
\hline
\end{tabular}
\end{table*}

Given the known temporal frame cycle and time resolution of the three FXT readout modes, pulsar targets with different spin periods were selected to verify and calibrate the in-flight timing system. As listed in Table\,\ref{tab:obs}, FF observed 1RXS J170849.0--400910 (\(\sim\!11\)~s, RXJ1708 for short hereafter), PW PSR~B1509--58 (\(150\)~ms), and TM the Crab pulsar (\(\sim\!33\)~ms) as well as the MSPs PSR~B1821--24 (\(3.05\)~ms) and PSR~B1937+21 (\(1.56\)~ms) due to their stable ephemerides and sharp pulse profiles. 

The in-flight timing calibration was performed during the performance verification and calibration (PV-CAL) phase from 2024 April 5 to July 9, with dedicated observations of these sources for each FXT readout mode. After the PV-CAL phase, the Crab pulsar is regularly monitored using TM mode. The in-flight observations for timing calibration up to mid-2025 are also listed in Table~\ref{tab:obs}. The ``exposure'' column denotes the effective exposure for FXTA and FXTB, respectively; a single value is shown when the two exposures are similar.

We processed the data of each observation (listed in the observation Identifier, ``obsID'', column in  Table\,\ref{tab:obs}) using FXT Data Analysis Software (\texttt{FXTDAS}, \citealt{2025RDTM....9..215Z}) along with FXT calibration database (\texttt{CALDB}, \citealt{2025RDTM....9..250L}), both in \texttt{Version 1.20}. The data reduction was carried out by the \texttt{fxtchain} pipeline with the source position listed in Table\,\ref{tab:srceph} as one group of the input parameters. We adopted the recommended Good Time Interval (GTI) selection criteria, 
\revt{
which include earth elevation angle greater than 5$^\circ$ (\texttt{ELV\textgreater{}5}), geomagnetic cut-off rigidity greater than 6\,GV (\texttt{COR\textgreater{}6}), outside the South Atlantic Anomaly region (\texttt{SAA==0}), and bright-earth elevation angle larger than 30$^\circ$ (\texttt{DYE\textunderscore{}ELV\textgreater{}30}). In addition, we only retained clean events with event grades in the range of 0--12 (\texttt{grade 0-12}), while rejecting bad and hot pixel events (\texttt{status==b0}).
}

The photon arrival times of FXT events are converted to the solar system barycenter (SSB) using the \texttt{fxtbary} tool in \texttt{FXTDAS} with DE430. An exception is the Crab pulsar, for which DE200 is used in order to refer to and compare with the radio ephemeris JBE.

\subsection{\revt{TOA Calculation}}

\subsubsection{\revt{Methodology}}

The phase of event with SSB time $t$ could be obtained given the pulsar ephemeris, 
\begin{equation}
    \phi = \phi_0 + \nu(t-t_0) + \frac{1}{2}\dot{\nu}(t-t_0)^2 + \frac{1}{6}\ddot{\nu}(t-t_0)^3 + \ldots,
\end{equation}
where $\phi_0$ is the phase at reference time $t_0$, and spin parameters $\nu$, $\dot{\nu}$ and $\ddot{\nu}$ are the corresponding frequency and its first- and second-order derivatives. 
The pulsar's TOA is determined by analyzing the accumulated pulse profile \citep{2022ApJS..259...14T},
\begin{equation}\label{eq:toa}
    \text{TOA} = t_0 + P\,\Delta\phi, 
\end{equation}
of which $P=1/\nu$ is the pulsar period, $\Delta\phi$ the phase of the reference point of the profile. In this work, the reference point is defined as the main peak of the pulse profile, with $t_0$ taken as the midpoint of each observation. 

\revt{
To determine the main peak phase $\Delta\phi$, a high statistical counting template profile is typically required. As will be detailed in the reference ephemeris for each pulsar, templates are either taken from references (for RXJ1708, PSR B1821--24, and PSR B1937+21) or constructed by stacking cross-correlated, phase-aligned FXT profiles (for PSR B1509--58 and the Crab pulsar). The TOA calculation is then performed as follows \citep{2019NatAs...3.1122G}: (1) For each FXT observation, obtain the reference time $t_0$ and the corresponding spin parameters $\nu$, $\dot{\nu}$ and $\ddot{\nu}$ at that time, then fold the event profile; (2) Shift the template profile to align with the folded profile through cross correlation; (3) Identify the maximum intensity phase in the shifted template, which is $\Delta\phi$; (4) Obtain the TOA using Equation\,\ref{eq:toa}.
}

\revt{
For the uncertainty calculation of each TOA, this work employs the approach that determined by the statistics and structure of the pulse profile \citep{2006JGCD...29...49S}, 
\begin{equation}
    \sigma_{\text{TOA}} = \frac{1}{2}W/{\text{SNR}},  
\end{equation}
In this equation, $W$ is the fraction of the pulse width, and 
\begin{equation}
    \text{SNR} = \frac{N_{S_{\text{pulsed}}}}{\sqrt{N_{B}+N_{S_{\text{nonpulsed}}}+N_{S_{\text{pulsed}}}}},
\end{equation}
denotes the signal-to-noise ratio (SNR) of the pulsed emission in the profile. Here, $N_{S_{\text{pulsed}}}$ is the pulsed source counts, while $N_{B}$ and $N_{S_{\text{nonpulsed}}}$ represent the background components and the non-pulsed source counts, respectively, within the pulse width range. As illustrated in the pulse profile plots of the five pulsars shown Section\,\ref{sec:results}, for RXJ1708 and PSR B1509--58 the entire phase from 0 to 1 is treated as the pulse range, where the total counts of $N_{B}+N_{S_{\text{nonpulsed}}}+N_{S_{\text{pulsed}}}$ are extracted; while the baseline $N_{B}+N_{S_{\text{nonpulsed}}}$ is determined by scaling the counts in the phase with minimum intensity to the entire phase range. For the double-peaked pulsars, the pulsed width is defined as the $\pm$0.05 phase range centered on the main peak position. The baseline is scaled from specific phase ranges to the pulse width range: 0.6--0.8 for the Crab pulsar, 0.2--0.4 for PSR B1821--24, and 0.7--0.9 for PSR B1937+21.
}

\begin{splitdeluxetable*}{llBcccccc}
\centering
\tablecaption{The pulsars and ephemerides used in this work \label{tab:srceph}}
\tablehead{
\colhead{\textbf{Source Name}}     & \colhead{\textbf{(RA, Dec) in J2000}}  &   \colhead{\textbf{$\nu$(Hz) }}        &  \colhead{\textbf{$\dot{\nu}$(Hz\,s$^{-1}$)} }   & \colhead{ \textbf{$\ddot{\nu}$(Hz\,s$^{-2}$)}}  &  \colhead{\textbf{Epoch(MJD) } } &\colhead{\textbf{ MJD range} }    & \colhead{\textbf{Reference} }
}
\colnumbers
\startdata 
\textbf{RXJ1708}       &  (17h08m46.87s, ~-$40^{\circ}08{'}52.44{''}$) \citep{2003ApJ...589L..93I}  &  0.0907878128     & $-1.676\times10^{-13}$        & ---                           & 60400         & 60200--60510  & \citealt{2026ApJ...999...85P}\\
\textbf{PSR B1509--58 }   & (15h13m55.640s, -$59^{\circ}08{'}09.2{''}$) ~\citep{2002ApJ...569..878G} & 6.56929922888     & $-6.5740275\times10^{-11}$    & $1.8536\times10^{-21}$        & 60200         & 59700--60800  & \citealt{2025ApJ...993..108Y} \\
\textbf{Crab pulsar }    & (05h34m31.972s, ~$22^{\circ}00{'}52.07{''}$)  (\href{http://www.jb.man.ac.uk/~pulsar/crab.html}{JBE})& --                & --                            & --                            & --            & --            & \href{http://www.jb.man.ac.uk/~pulsar/crab.html}{JBE} \\
\textbf{PSR B1821--24 }   & (18h24m32.008s, -$24^{\circ}52{'}10.85{''}$) \citep{2019MNRAS.490.4666P} & 327.40554427917   & $-1.73510\times10^{-13}$      & $-2.1\times10^{-25}$          & 57920         &  57956--60113          & \citealt{2024Univ...10..174Z} \\
\textbf{PSR B1937+21}    & (19h39m38.561s, ~$21^{\circ}34{'}59.13{''}$) \citep{2019MNRAS.490.4666P} & 641.92822360938   &  $-4.33064\times10^{-14}$     & $-1.8\times10^{-26}$          & 57920         & 57945--60134            & \citealt{2024Univ...10..174Z} \\
\enddata  
\end{splitdeluxetable*}

\subsubsection{\revt{Reference Ephemerides of Pulsars}}

The ephemerides referenced in this work are listed in Table\,\ref{tab:srceph}. RXJ1708 is an anomalous X-ray pulsar with frequent historical glitches (\citealt{2026ApJ...999...85P}  and references therein). We adopt the timing solution over the MJD range 60200--60510 \revt{and the template pulse profile} derived from NICER data \citep{2026ApJ...999...85P}, which encompasses the date of FXT observations. For FXT events, we choose celestial photons falling within the 2$^\prime$ radius circle centered on the source position in the energy range of 0.7--10\,keV. This energy band  aligns with that used in \citet{2026ApJ...999...85P}.

PSR B1509--58 is a young, isolated, and bright soft gamma-ray pulsar, which is surrounded by its pulsar wind nebula and its host the supernova remnant G320.4--1.2 \citep{1982ApJ...256L..45S,1982ApJ...262L..31M}. In this work, we adopt the timing model derived by \citealt{2025ApJ...993..108Y} using Fermi-LAT 30--1000\,MeV data in the MJD range 59700--60800, which covers the FXT observation date. FXT data are selected in the energy range of 0.5--10\,keV within the 1$^{\prime}$ radius circle around the source. 
\revt{
And the template pulse profile is built by accumulating phase-aligned profiles with FXT data through cross-correlation. 
}
To verify the time system of FXT in PW mode, the FXT timing results are compared with those of NICER. Clean NICER data in the same energy range, collected in 2024 from March 27 to May 24 (obsIDs 7020020103--7020020112), were employed for this analysis. In this work, the \texttt{barycorr} tool in \texttt{HEASOFT FTOOL} package is used for the barycentric correction of NICER data.

Although the Crab pulsar exhibits variable timing parameters, it remains a widely used source for timing calibration of space-based instruments spanning the X-ray to gamma-ray bands, owing to its high luminosity, broadband emission, and extensive monitoring history. In this work, we employ the JBE radio ephemeris for the Crab pulsar; consequently, DE200 is adopted for barycentric time correction. Events in the 0.5--10 keV energy range in all TM readout pixels are selected. 
\revt{
The template pulse profile is also obtained by adding cross-correlated, phase-aligned FXT profiles.}  Because the main peak position of the Crab pulsar varies with energy \citep{2010ApJ...708..403M,2022ApJS..259...14T}, we also analyzed clean NICER data since 2025 January 20 (obsIDs 6013010125--8013010101) in the the same 0.5--10 keV band for comparison. 
\revtt{For observations with empty GTIs in clean data, the calibrated unfiltered (\texttt{ufa}) data were utilized for timing analysis when the exposure exceeded 100\,s. }

PSR B1821-24 and PSR B1937+21 are MSPs with long-term timing stability. We adopt the timing solutions from \citet{2024Univ...10..174Z} \revt{and the pulse profiles there as the templates} , which were derived through six years of NICER observations up to 2023 September. For PSR B1821--24, the data from FXTA RAWX columns [190,194] are selected, while for FXTB columns [189,192] are chosen. In case of PSR B1937+21, events within RAWX columns [188,192] selected for FXTA and [188,191] for FXTB. To keep consistency with the reference ephemerides, FXT data were selected in the 1.0--5.5\,keV energy band for PSR B1821--24 and 1.15--5.55\,keV for PSR B1937+21.

\section{Timing Results} \label{sec:results}

\begin{figure*}[]
    \centering
    \includegraphics[width=0.45\linewidth]{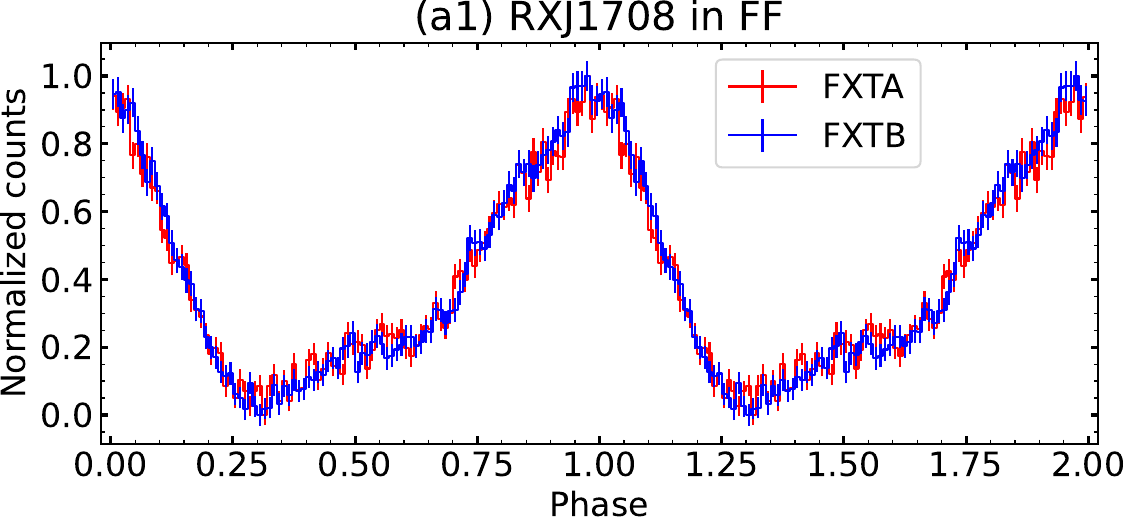}
    \includegraphics[width=0.45\linewidth]{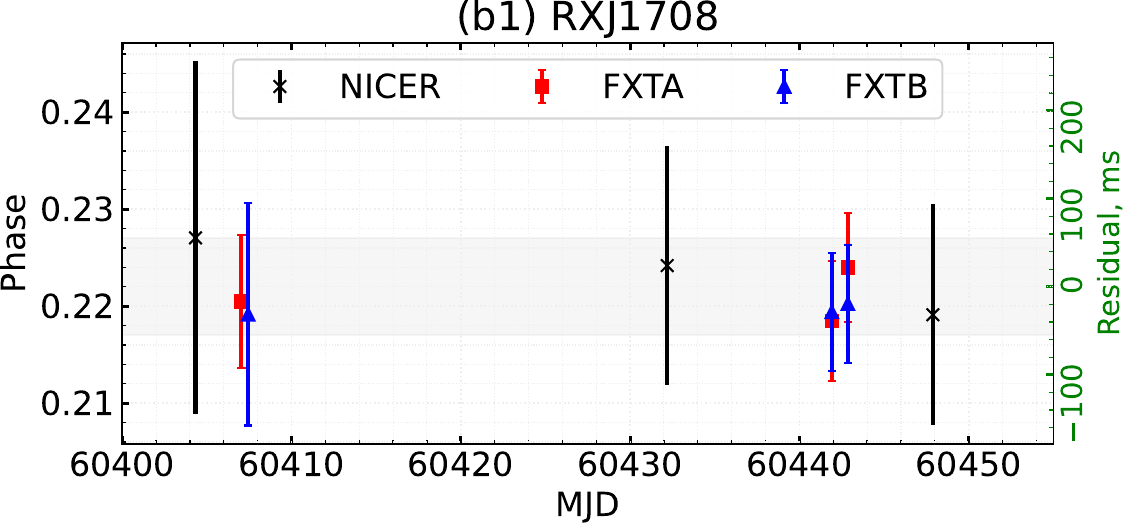}
    \includegraphics[width=0.45\linewidth]{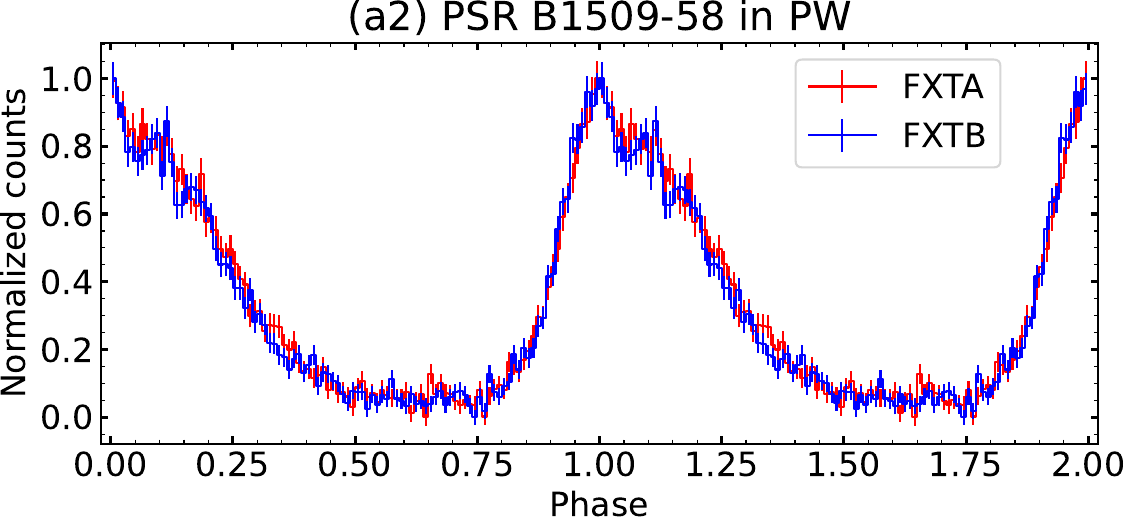}
    \includegraphics[width=0.45\linewidth]{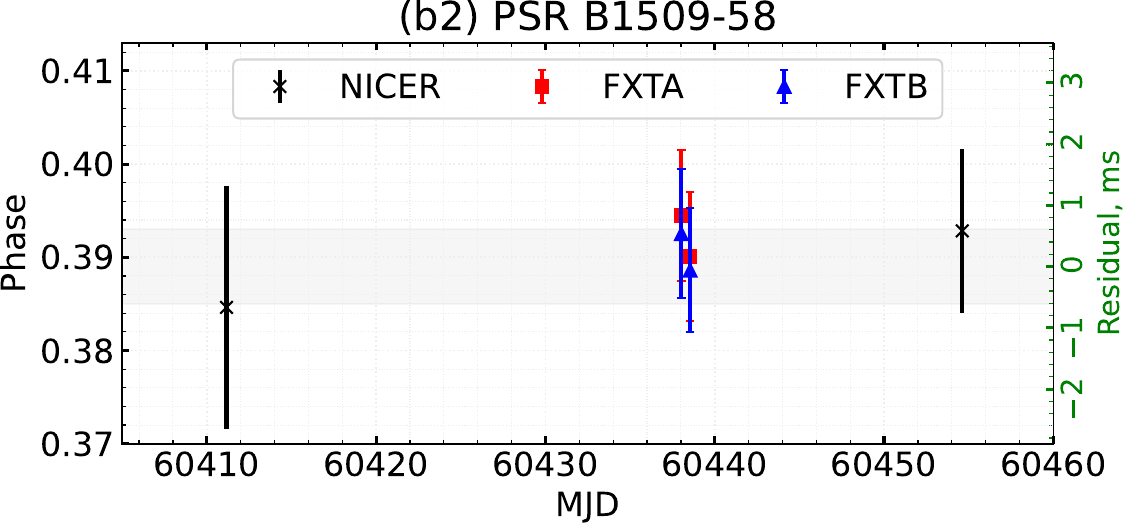}
    \includegraphics[width=0.45\linewidth]{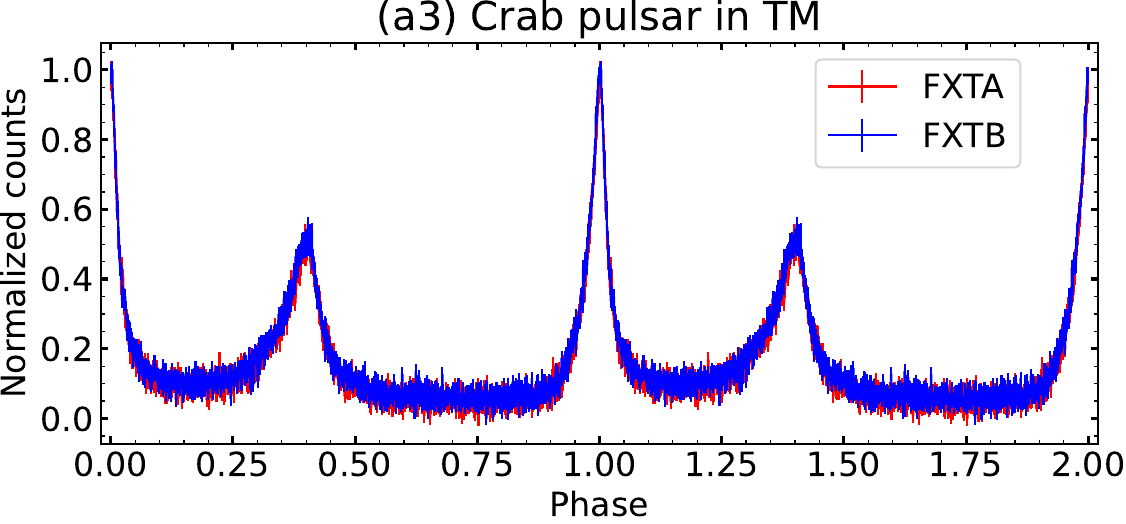}
    \includegraphics[width=0.45\linewidth]{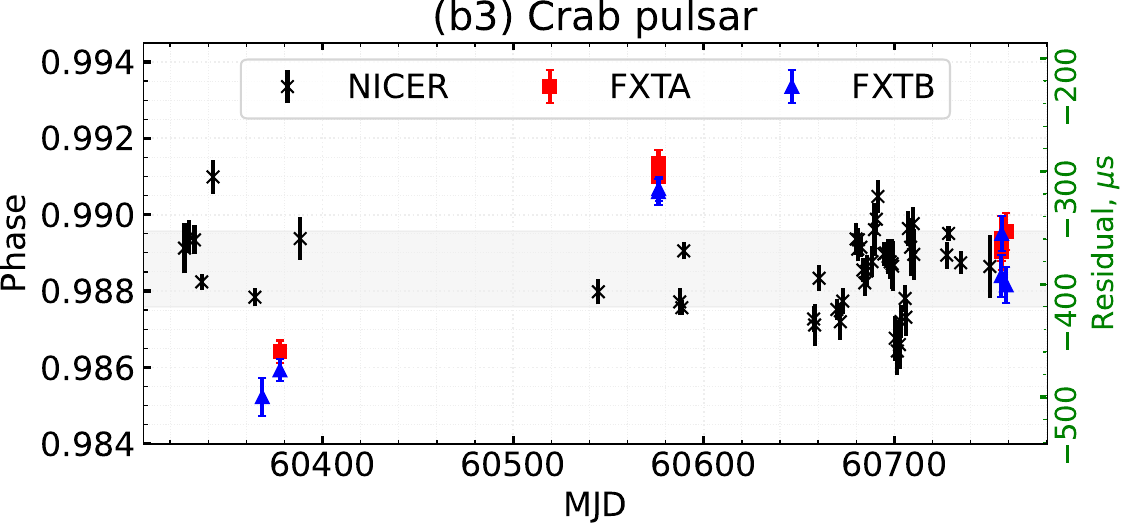}
    \includegraphics[width=0.45\linewidth]{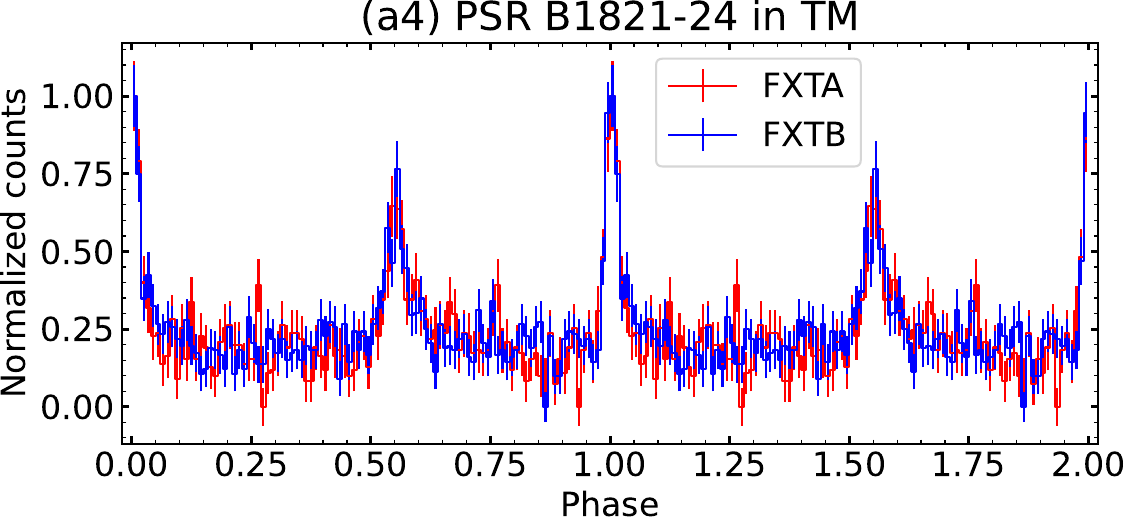}
    \includegraphics[width=0.45\linewidth]{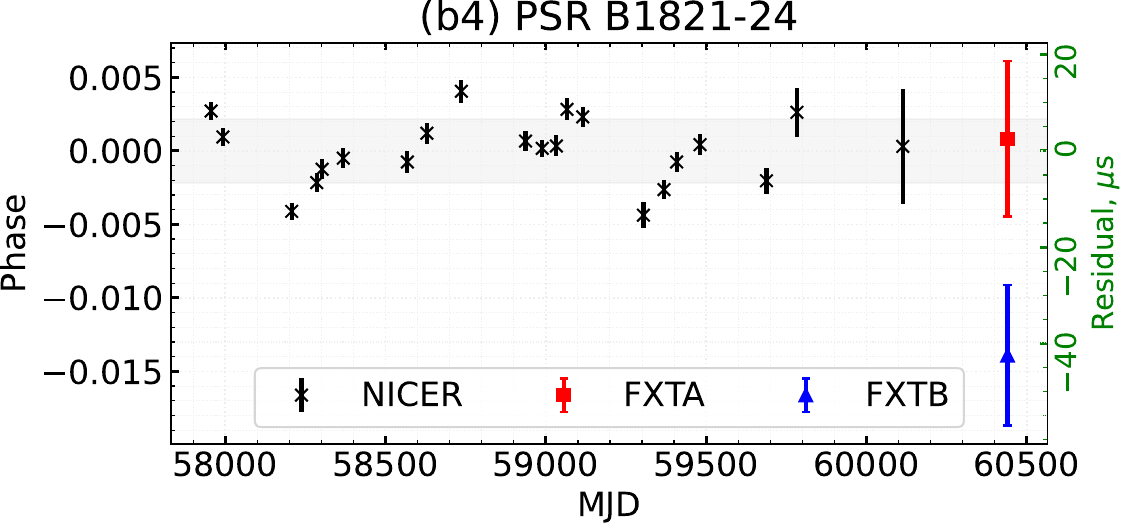}
    \includegraphics[width=0.45\linewidth]{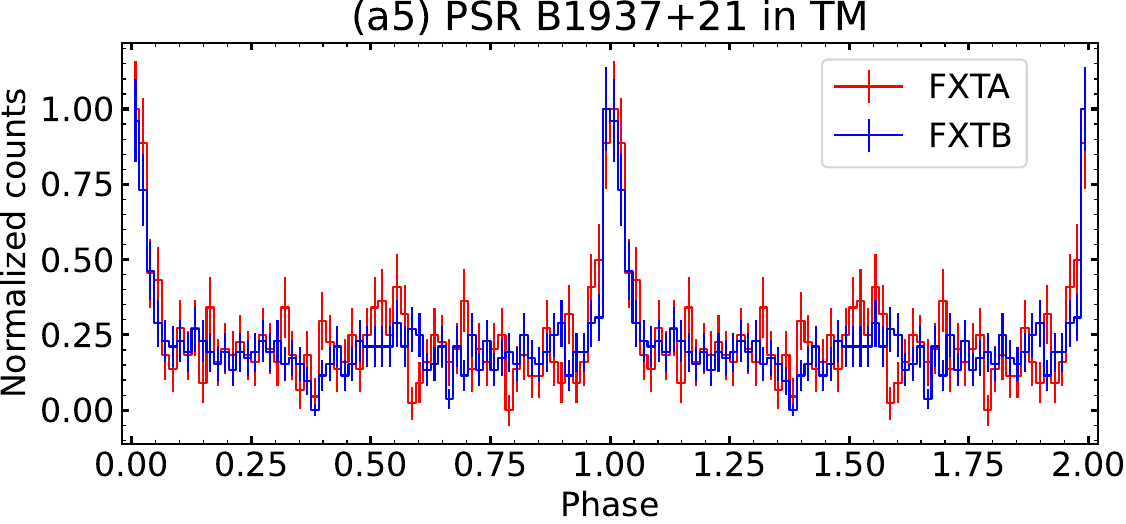}
    \includegraphics[width=0.45\linewidth]{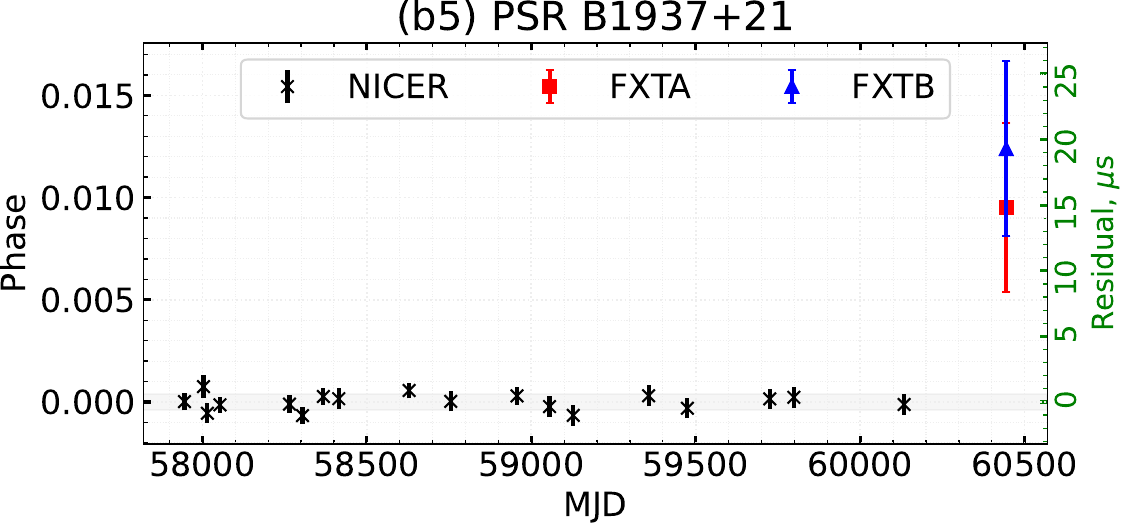}    
    \caption{The pulse profiles (left panels) and TOA phases (right panels) obtained with FXTA (in red) and FXTB (in blue) data for the five pulsars, i.e., RXJ1708 in FF mode, PSR B1509--58 in PW mode, and Crab pulsar, PSR B1821--24 and PSR B1937+21 in TM mode.  Main peak positions from reference ephemerides (shaded areas) and NICER TOAs (black crosses) are also plotted in comparison. 
    \revt{The secondary y-axes of TOA panels denote time residuals relative to the ephemeris TOA.}
    }
    \label{fig:5pulsar}
\end{figure*}

\begin{table*}[htb]
\color{mycolor}
\centering
\caption{\revt{Summary of the comparison and TOA residuals of FXT relative to the reference ephemerides}\label{tab:summary_residual}}
\resizebox{\linewidth}{!}{
\begin{tabular}{c|c|c|c|c|c|c|c}
\hline
\hline
\multirow{2}{*}{\textbf{Mode}} & \multirow{2}{*}{\textbf{Source}} & \multicolumn{3}{c|}{\textbf{Mean Phase}} & \multicolumn{2}{c|}{\textbf{Residual}} & \textbf{Systematic Delay} \\ 
\cline{3-7}
    {}                         &                                  &  \textbf{FXTA}    &      \textbf{FXTB}    & \textbf{Reference Phase}          &  \textbf{FXTA}    &      \textbf{FXTB}    &  \textbf{FXTA - FXTB} \\
\hline
\textbf{FF}                    &  \text{RXJ1708}                       &  $0.221^{\pm0.006(stat)}_{\pm0.002(sys)}$      &  $0.220^{\pm0.008(stat)}_{\pm0.001(sys)}$          &    $0.222\pm0.005$             &     -11$\pm$91\,ms              & -26$\pm$103\,ms    &  $15^{\pm111(stat)}_{\pm21(sys)}$\,ms\\
\hline
\textbf{PW}     &  \text{PSR B1509--58}   & $0.392^{\pm0.007(stat)}_{\pm0.002(sys)}$       &  $0.391^{\pm0.007(stat)}_{\pm0.002(sys)}$        &    $0.389^{\pm0.011(stat)}_{\pm0.004(sys)}$              & 0.54$\pm$2.10\,ms   & 0.28$\pm$ 2.08\,ms & $0.26^{\pm1.48(stat)}_{\pm0.03(sys)}$\,ms \\
\hline
\multirow{4}{*}{\textbf{TM}}                
&  \multirow{2}{*}{\text{Crab pulsar}} &   \multirow{2}{*}{$0.9895^{\pm0.0004(stat)}_{\pm0.0016(sys)}$}     &  \multirow{2}{*}{$0.9884^{\pm0.0004(stat)}_{\pm0.0020(sys)}$}            &    $^{*}$1.      & 
$-357^{\pm14(stat)}_{\pm54(sys)}$\,\us &   $-394^{\pm14(stat)}_{\pm67(sys)}$\,\us  & \multirow{4}{*}{$20^{\pm19(stat)}_{\pm21(sys)}$\,\us}\\   
\cline{5-7}
&     &    &    &    $^{\dagger}$$0.9886^{\pm0.0004(stat)}_{\pm0.0010(sys)}$              & $30\pm67$\,\us &   $-7\pm78$\,\us  & \\        
\cline{2-7}
&  \text{PSR B1821--24}   & $0.0008^{\pm0.0053(stat)}$ & $-0.0139^{\pm0.0048(stat)}$ & $0^{\pm0.0009(stat)}_{\pm0.0022(sys)}$                  &  $3\pm18$\,\us                    & $-42\pm16$\,\us & \\    
\cline{2-7}
&  \text{PSR B1937+21}    & $0.0095^{\pm0.0041(stat)}$ & $0.0124^{\pm0.0043(stat)}$ & $0^{\pm0.0005(sta)}_{\pm0.0004(sys)}$ &  $15\pm7$\,\us                   &    $19\pm7$\,\us  &  \\      

\hline
\hline
\end{tabular}
} 
\footnotesize
\tablecomments{\revt{For the Crab pulsar, * row is relative to JBE, and $^{\dagger}$ to the mean phase of NICER TOA data.}}
\end{table*}

The timing results of the five pulsars analysed using FXTA and FXTB data are presented in Figure\,\ref{fig:5pulsar}. The left panels show the pulse profiles with FXTA in red and FXTB in blue. While the right panels display the TOAs with FXTA in red squares and FXTB in blue triangles. Also overlaid for comparison are the reference ephemerides, \revt{i.e.} main peak position regions in shaded area and TOAs in black crosses. 
\revt{
Table\,\ref{tab:summary_residual} summarizes these comparison results and TOA residuals in FXTA and FXTB relative to the reference ephemerides. The mean phase is calculated by averaging all measured TOA phases, with the statistical error of the averaged TOA defined as the mean of individual TOA statistical uncertainties. The systematic error is determined as the standard deviation of the TOA values. TOA residuals are obtained relative to the TOA phase in the reference ephemerides, and the residual error is calculated through error propagation. 
Also presented in this table is the systematic delay between the residuals of FXTA and FXTB. It is calculated by averaging the residual delays of FXTA relative to FXTB over observations where TOAs from both modules were available. The statistical and systematic uncertainties of this mean delay are estimated using the same procedure as for the averaged TOA uncertainties.
}


In Figure\,\ref{fig:5pulsar}(b1), the TOAs of RXJ1708 obtained from FXT data in FF mode are compared with the reference main peak position of $0.222\pm0.005$ (shaded region) and the NICER TOAs (black crosses) from \citet{2026ApJ...999...85P}. Given obsID 05100000001 has a long exposure, the FXT data were split into two intervals containing equal event counts. The resulting averaged TOA phase is $0.221{^{\pm\revt{0.006(stat)}}_{\pm\revt{0.002(sys)}}}$ with FXTA and $0.220^{\pm\revt{0.008(stat)}}_{\pm0.001(sys)}$ in FXTB.
\revt{
Correspondingly, as given in Table\,\ref{tab:summary_residual} the TOA residuals relative to the reference ephemeris are -11$\pm$91\,ms in FXTA and -26$\pm$103\,ms in FXTB, respectively. Currently the statistical uncertainty dominates the residual error. In addition, the systematic delay between FXTA and FXTB is obtained as $15^{\pm111(stat)}_{\pm21(sys)}$\,ms. Therefore, it could be concluded that the timing system of FXTA and FXTB in FF mode are consistent with each other, and that they both are aligned well with NICER results.
}


The 2.2\,ms frame interval of PW mode corresponds to 0.015 phase of the 150\,ms spin period of the calibration reference source PSR B1509--58. The timing results in PW mode of FXTA and FXTB are plotted in comparison with NICER results in Figure\,\ref{fig:5pulsar}(b2). FXT data were split into two groups with equal event numbers. The NICER data, results of which are shown in black crosses, are also grouped before (obsIDs from 7020020103 to 7020020111) and after (obsID 7020020112) the FXT observations. The average TOA phase of FXTA is $0.392^{\pm\revt{0.007(stat)}}_{\pm0.002(sys)}$  and FXTB $0.391^{\pm\revt{0.007(stat)}}_{\pm0.002(sys)}$. The NICER result is $0.389^{\pm\revt{0.011(stat)}}_{\pm0.004(sys)}$, with this systematic error represented by grey-shaded region in this panel. 
\revt{
The TOA residuals relative to NICER is 0.54$\pm$2.10\,ms for FXTA and 0.28$\pm$ 2.08\,ms for FXTB. And the systematic residual between FXTA and FXTB in PW mode is $0.26^{\pm1.48(stat)}_{\pm0.03(sys)}$\,ms.
}

The TOA residuals of Crab pulsar using FXT data in TM mode relative to the radio ephemeris are compared with the NICER observations in Figure\,\ref{fig:5pulsar}(b3). Each data point on this panel signifies a single observation. Table\,\ref{tab:summary_residual} presents the relative shift of the main peak to JBE. The mean advance measured by FXTA is \revt{$357^{\pm14(stat)}_{\pm54(sys)}$}\,\us, by FXTB  is $394^{\pm\revt{14(stat)}}_{\pm\revt{67(sys)}}$\,\us. Contemporary observations of NICER from 2024 January mid to 2025 March presents an advance of \revt{$387^{\pm15(stat)}_{\pm34(sys)}$}\,\us, with the systematic error shown in shaded region. 
\revt{
The TOA residuals relative to the mean phase of NICER TOAs is $30\pm67$\,\us in FXTA and $-7\pm78$\,\us in FXTB. 
} \revtt{As shown in Table\,\ref{tab:summary_residual}, the uncertainties here are predominantly systematic. To mitigate the impact of timing noise from the Crab pulsar, we restrict the analysis to FXTA, FXTB, and NICER TOAs separated by no more than 10 days. This limited the observations to MJD 60750--60759, resulting in offsets of $23^{\pm33(stat)}_{\pm7(sys)}$\,\us for FXTA and $14^{\pm33(stat)}_{\pm20(sys)}$\,\us for FXTB  relative to NICER.}
 

The FXT TOA residuals of PSR B1821--24 and PSR B1937+21 in TM mode are shown in comparison with the reference ephemerides of \citet{2024Univ...10..174Z} in Figure\,\ref{fig:5pulsar}(b4) and Figure\,\ref{fig:5pulsar}(b5). 
For PSR B1821--24, the twice FXT observations are combined together; the TOA residual obtained with FXTA is $3\pm\revt{18}$\,\us, and FXTB $-42\pm\revt{16}$\,\us. For PSR B1937+21, FXTA shows a delay of 15$\pm$\revt{7}\,\us, and FXTB 19$\pm$\revt{7}\,\us. 

\revt{
The systematic uncertainties of TOA residuals derived for FXTA and FXTB provide an initial estimate of the in-flight absolute timing accuracy. As presented in Table\,\ref{tab:summary_residual}, the systematic uncertainty based on the Crab pulsar is 54\,\us for FXTA and 67\,\us for FXTB, which could be treated as the upper limit  of the timing accuracy considering the strong timing noise associated with this pulsar. Based on the limited observations of the two more stable pulsars PSR B1821--24 and PSR B1937+21, the combination of their residuals yields a preliminary timing accuracy of 6\,\us for FXTA and 31\,\us for FXTB.
}

\begin{figure}[b]
    \centering
    \includegraphics[width=0.9\linewidth]{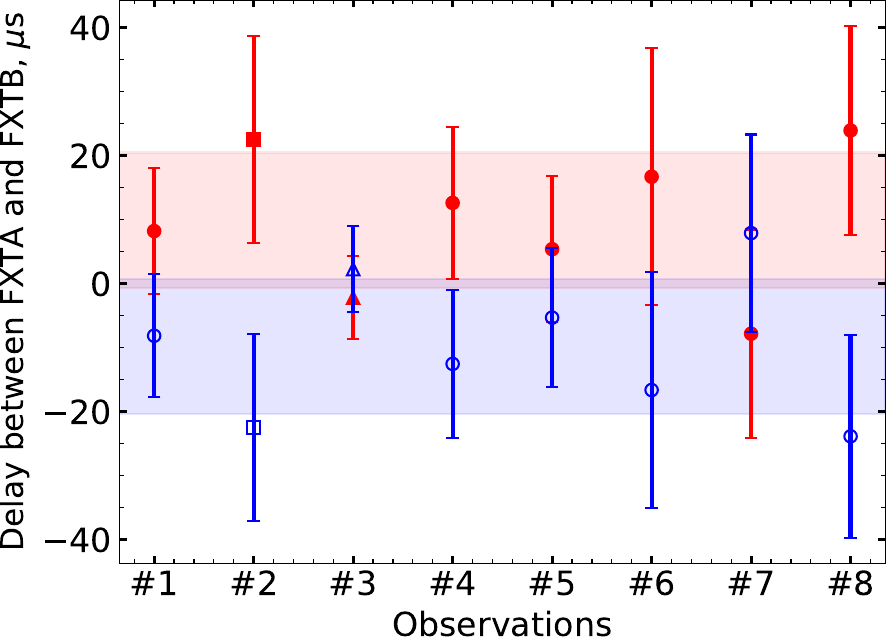}
    \caption{\revt{The systematic delay between FXTA (red solid points) and FXTB (blue hollow points) in TM mode. Data represent measurements from the Crab pulsar (dotted points), PSR B1821--24 (square points), and PSR B1937+21 (triangle points), with the mean residual of FXTA and FXTB at each observation normalized to zero. The red (blue) shaded region represents the systematic uncertainty range of the averaged relative delay of FXTA (FXTB).}}
    \label{fig:ABdelay}
\end{figure}

\revt{The systematic delay between the FXTA and FXTB time results in TM mode is determined using TOA residuals from observations of the Crab pulsar, PSR B1821--24, and PSR B1937+21. Figure\,\ref{fig:ABdelay} shows the relative residuals of FXTA and FXTB, with the mean of each observation normalized to zero. Based on these results, the systematic delay of FXTA relative to FXTB is found to be $20^{\pm19(stat)}_{\pm21(sys)}$\,\us.
}

\section{Conclusions} \label{sec:con}

The in-flight timing performance for each of the three readout modes (FF, PW, and TM) of the two EP/FXT modules, FXTA and FXTB, was verified and calibrated through pulsar observations. 
\revt{The timing results of RXJ1708 in FF mode and PSR B1509--58 in PW mode demonstrate that the timing systems of FXTA and FXTB in these two mode are mutually consistent and both aligned well with the NICER timing results.}

\revt{
In TM mode, timing results of the Crab pulsar, PSR B1821--24, and PSR B1937+21 also show good alignment with NICER within uncertainties. The measured advance of the main peak of the Crab pulsar is $357^{\pm14(stat)}_{\pm54(sys)}$\,\us in FXTA and $394^{\pm14(stat)}_{\pm67(sys)}$\,\us. in FXTB, relative to the radio ephemeris JBE. 
The systematic uncertainty of the Crab pulsar residuals set an upper limit for the in-flight timing accuracy of 54\,\us for FXTA and 67\,\us for FXTB. Current observations of PSR B1821--24 and PSR B1937+21 yields an improved estimate of 6\,\us for FXTA and 31\,\us for FXTB, respectively. Based on the residuals of these three pulsars, the systematic delay of the time system of FXTA relative to FXTB is found to be $20^{\pm19(stat)}_{\pm21(sys)}$\,\us.}
Regular TM-mode observations on pulsars are scheduled several times per year to monitor the in-flight timing performance of FXT and to establish a reliable absolute timing accuracy by accumulating additional data.

\begin{acknowledgments}
This work is supported by the National Natural Science Foundation of China (Grant No.\,42327802), the Strategic Priority Program on Space Science of Chinese Academy of Sciences (Grants No.\,E02212A02S, No.\,XDA15310103), and National Key R\&D Program of China No.\,2025YFF0511104. ZJ\&FLM acknowledge the National Space Science Data Center of China (\url{https://www.nssdc.ac.cn}) for providing data resources/computing services/data analysis environment.
\end{acknowledgments}

\bibliography{fxt_inflight_timecal}{}

\begin{thebibliography}{}
\expandafter\ifx\csname natexlab\endcsname\relax\def\natexlab#1{#1}\fi
\providecommand{\url}[1]{\href{#1}{#1}}
\providecommand{\dodoi}[1]{doi:~\href{http://doi.org/#1}{\nolinkurl{#1}}}
\providecommand{\doeprint}[1]{\href{http://ascl.net/#1}{\nolinkurl{http://ascl.net/#1}}}
\providecommand{\doarXiv}[1]{\href{https://arxiv.org/abs/#1}{\nolinkurl{https://arxiv.org/abs/#1}}}

\bibitem[{{Bachetti} {et~al.}(2026){Bachetti}, {Terada}, {Shidatsu},
  {Markwardt}, {Chen}, {Cui}, {Cusumano}, {Han}, {Jia}, {Kang}, {Kashyap},
  {Kuiper}, {Li}, {Motogami}, {Ota}, {Pagliarella}, {Pottschmidt}, {Rosen},
  {Rots}, {Sawada}, {Sugizaki}, {Takagi}, {Takahashi}, {Tamagawa}, {Tuo},
  {Yang}, {Yoshimoto}, \& {Zhang}}]{2026PASP..138a4501B}
{Bachetti}, M., {Terada}, Y., {Shidatsu}, M., {et~al.} 2026, \pasp, 138,
  014501, \dodoi{10.1088/1538-3873/ae2b36}

\bibitem[{{Bogdanov}(2013)}]{2013ApJ...762...96B}
{Bogdanov}, S. 2013, \apj, 762, 96, \dodoi{10.1088/0004-637X/762/2/96}

\bibitem[{{Chen} {et~al.}(2025){Chen}, {Wang}, {Yang}, {Zhao}, {Ma}, {Wang},
  {Han}, {Cui}, {Ban}, {Bi}, {Bianucci}, {Burwitz}, {Cai}, {Cai}, {Chen},
  {Chen}, {Chen}, {Cheng}, {Ding}, {Eder}, {Feng}, {Feng}, {Friedrich}, {Geng},
  {Gao}, {Gao}, {Guan}, {Hou}, {Huo}, {Jia}, {Keereman}, {Li}, {Li}, {Li},
  {Li}, {Li}, {Li}, {Li}, {Liao}, {Liu}, {Liu}, {Liu}, {Lu}, {Luo},
  {Meidinger}, {Qiang}, {Santovincenzo}, {Sheng}, {Sun}, {Tang}, {Tian},
  {Valsecchi}, {Vernani}, {Wang}, {Wang}, {Wang}, {Wang}, {Wang}, {Wang},
  {Xie}, {Xu}, {Xue}, {Yan}, {Yang}, {Yu}, {Yuan}, {Zhang}, {Zhang}, {Zhang},
  {Zhang}, {Zhang}, {Zhang}, {Zhang}, {Zhao}, {Zhao}, \&
  {Zhu}}]{2025RDTM....9..198C}
{Chen}, Y., {Wang}, J., {Yang}, Y., {et~al.} 2025, Radiation Detection
  Technology and Methods, 9, 198, \dodoi{10.1007/s41605-025-00558-0}

\bibitem[{{Cheng} {et~al.}(2024){Cheng}, {Ling}, {Zhang}, {Sun}, {Sun}, {Liu},
  {Dai}, {Jia}, {Pan}, {Wang}, {Zhao}, {Chen}, {Cheng}, {Fu}, {Han}, {Li},
  {Li}, {Ma}, {Xue}, {Yan}, {Zhang}, {Wang}, {Yang}, {Zhao}, \&
  {Yuan}}]{2024ExA....57...10C}
{Cheng}, H., {Ling}, Z., {Zhang}, C., {et~al.} 2024, Experimental Astronomy,
  57, 10, \dodoi{10.1007/s10686-024-09932-0}

\bibitem[{{Cheng} {et~al.}(2025){Cheng}, {Zhao}, {Tao}, {Feng}, {Coti Zelati},
  {Pan}, {Wang}, {Wang}, {Ge}, {Rau}, {Marino}, {Zhang}, {Zhang}, {Carotenuto},
  {Ji}, {Jin}, {Li}, {Liu}, {Liu}, {Qiao}, {Rea}, {Soria}, {Wang}, {Yan},
  {Yuan}, {Zhang}, {Zhang}, {Zhang}, {Zhang}, {Beardmore}, {Bright}, {Chen},
  {Fan}, {Fu}, {Fynbo}, {Hu}, {Jin}, {Jonker}, {Kong}, {Kuulkers}, {Li}, {Li},
  {Lin}, {Liu}, {Liu}, {Liu}, {Liu}, {Lu}, {Maitra}, {Mu}, {Ng}, {Qiu},
  {Tinyanont}, {Wang}, {Wen}, {Weng}, {Wu}, {Xu}, {Yan}, {Yang}, {Zhang},
  {Zhang}, {Zhao}, {Cai}, {Chen}, {Chen}, {Cui}, {Cui}, {Hu}, {Huang}, {Jia},
  {Jin}, {Ling}, {Liu}, {Sun}, {Sun}, {Xu}, {Zhang}, {Zhang}, \&
  {Zhang}}]{2025ApJ...991L..41C}
{Cheng}, H.~Q., {Zhao}, Q.~C., {Tao}, L., {et~al.} 2025, \apjl, 991, L41,
  \dodoi{10.3847/2041-8213/adf104}

\bibitem[{{Cui} {et~al.}(2023){Cui}, {Wang}, {Zhao}, {Zhang}, {Meidinger},
  {Yang}, {Keil}, {Zhang}, {Huo}, {Wang}, {Song}, {Lu}, {Ma}, {Wang}, {Xu},
  {Zhu}, {Li}, {Li}, {Luo}, {Han}, {Zhao}, {Hou}, {Yang}, {Geng}, {Li}, {Chen},
  {Tang}, {Chen}, \& {Chen}}]{2023ExA....55..603C}
{Cui}, W., {Wang}, H., {Zhao}, X., {et~al.} 2023, Experimental Astronomy, 55,
  603, \dodoi{10.1007/s10686-023-09891-y}

\bibitem[{{Cusumano} {et~al.}(2012){Cusumano}, {La Parola}, {Capalbi}, {Perri},
  {Beardmore}, {Burrows}, {Campana}, {Kennea}, {Osborne}, {Sbarufatti}, \&
  {Tagliaferri}}]{2012A&A...548A..28C}
{Cusumano}, G., {La Parola}, V., {Capalbi}, M., {et~al.} 2012, \aap, 548, A28,
  \dodoi{10.1051/0004-6361/201219968}

\bibitem[{{Gaensler} {et~al.}(2002){Gaensler}, {Arons}, {Kaspi}, {Pivovaroff},
  {Kawai}, \& {Tamura}}]{2002ApJ...569..878G}
{Gaensler}, B.~M., {Arons}, J., {Kaspi}, V.~M., {et~al.} 2002, \apj, 569, 878,
  \dodoi{10.1086/339354}

\bibitem[{{Ge} {et~al.}(2019){Ge}, {Lu}, {Yan}, {Weng}, {Zhang}, {Wang},
  {Wang}, {Li}, \& {Zhang}}]{2019NatAs...3.1122G}
{Ge}, M.~Y., {Lu}, F.~J., {Yan}, L.~L., {et~al.} 2019, Nature Astronomy, 3,
  1122, \dodoi{10.1038/s41550-019-0853-5}

\bibitem[{{Israel} {et~al.}(2003){Israel}, {Covino}, {Perna}, {Mignani},
  {Stella}, {Campana}, {Marconi}, {Bono}, {Mereghetti}, {Motch}, {Negueruela},
  {Oosterbroek}, \& {Angelini}}]{2003ApJ...589L..93I}
{Israel}, G.~L., {Covino}, S., {Perna}, R., {et~al.} 2003, \apjl, 589, L93,
  \dodoi{10.1086/375832}

\bibitem[{{Jin}(2021)}]{2021ApJ...920...60J}
{Jin}, C. 2021, \apj, 920, 60, \dodoi{10.3847/1538-4357/ac1674}

\bibitem[{{Li} {et~al.}(2025{\natexlab{a}}){Li}, {Jia}, {Song}, {Chen}, {Zhao},
  {Zhang}, {Guan}, {Zhao}, {Han}, {Cui}, {Xu}, \& {Wang}}]{2025RDTM....9..250L}
{Li}, C., {Jia}, S.-M., {Song}, L.-M., {et~al.} 2025{\natexlab{a}}, Radiation
  Detection Technology and Methods, 9, 250, \dodoi{10.1007/s41605-024-00512-6}

\bibitem[{{Li} {et~al.}(2021){Li}, {Lin}, {Xiong}, {Ge}, {Li}, {Li}, {Lu},
  {Zhang}, {Tuo}, {Nang}, {Zhang}, {Xiao}, {Chen}, {Song}, {Xu}, {Liu}, {Jia},
  {Cao}, {Qu}, {Zhang}, {Gu}, {Liao}, {Zhao}, {Tan}, {Nie}, {Zhao}, {Zheng},
  {Zheng}, {Luo}, {Cai}, {Li}, {Xue}, {Bu}, {Chang}, {Chen}, {Chen}, {Chen},
  {Chen}, {Chen}, {Cui}, {Cui}, {Deng}, {Dong}, {Du}, {Fu}, {Gao}, {Gao},
  {Gao}, {Gu}, {Guan}, {Guo}, {Han}, {Huang}, {Huo}, {Jiang}, {Jiang}, {Jin},
  {Jin}, {Kong}, {Li}, {Li}, {Li}, {Li}, {Li}, {Li}, {Li}, {Liang}, {Liu},
  {Liu}, {Liu}, {Liu}, {Liu}, {Lu}, {Lu}, {Luo}, {Ma}, {Meng}, {Ou}, {Sai},
  {Shang}, {Song}, {Sun}, {Tao}, {Wang}, {Wang}, {Wang}, {Wang}, {Wang}, {Wen},
  {Wu}, {Wu}, {Wu}, {Xiao}, {Xu}, {Yang}, {Yang}, {Yang}, {Yang}, {Yi}, {Yin},
  {You}, {Zhang}, {Zhang}, {Zhang}, {Zhang}, {Zhang}, {Zhang}, {Zhang},
  {Zhang}, {Zhang}, {Zhang}, {Zhang}, {Zhang}, {Zhang}, {Zhang}, {Zhang},
  {Zhang}, {Zhou}, {Zhou}, {Zhu}, {Zhu}, \& {Zhuang}}]{2021NatAs...5..378L}
{Li}, C.~K., {Lin}, L., {Xiong}, S.~L., {et~al.} 2021, Nature Astronomy, 5,
  378, \dodoi{10.1038/s41550-021-01302-6}

\bibitem[{{Li} {et~al.}(2022){Li}, {Ge}, {Lin}, {Zhang}, {Song}, {Cao},
  {Zhang}, {Lu}, {Xu}, {Xiong}, {Tuo}, {Tan}, {Jiang}, {Qu}, {Zhang}, {Wang},
  {Wang}, {Zhang}, {Zhang}, {Li}, {Liu}, {Li}, {Bu}, {Cai}, {Chen}, {Chen},
  {Chang}, {Chen}, {Chen}, {Chen}, {Cui}, {Du}, {Gao}, {Gao}, {Gu}, {Guan},
  {Guo}, {Han}, {Huang}, {Huo}, {Jia}, {Jin}, {Kong}, {Li}, {Li}, {Li}, {Li},
  {Li}, {Li}, {Liang}, {Liao}, {Liu}, {Liu}, {Liu}, {Lu}, {Luo}, {Luo}, {Ma},
  {Ma}, {Ma}, {Meng}, {Nang}, {Nie}, {Ou}, {Ren}, {Sai}, {Song}, {Sun}, {Tao},
  {Wang}, {Wang}, {Wang}, {Wang}, {Wen}, {Wu}, {Wu}, {Wu}, {Xiao}, {Yang},
  {Yang}, {Yi}, {Yin}, {You}, {Yu}, {Zhang}, {Zhang}, {Zhang}, {Zhang},
  {Zhang}, {Zhang}, {Zhang}, {Zhao}, {Zhao}, {Zheng}, \&
  {Zhou}}]{2022ApJ...931...56L}
{Li}, X., {Ge}, M., {Lin}, L., {et~al.} 2022, \apj, 931, 56,
  \dodoi{10.3847/1538-4357/ac6587}

\bibitem[{{Li} {et~al.}(2025{\natexlab{b}}){Li}, {Kuiper}, {Pan}, {Xu}, {Chen},
  {Ge}, {Huang}, {Jia}, {Li}, {Song}, {Qu}, {Zhang}, {Tao}, {Feng}, {Zhang}, \&
  {Falanga}}]{2025ApJ...990...15L}
{Li}, Z., {Kuiper}, L., {Pan}, Y., {et~al.} 2025{\natexlab{b}}, \apj, 990, 15,
  \dodoi{10.3847/1538-4357/adf0ff}

\bibitem[{{Lyne} {et~al.}(1993){Lyne}, {Pritchard}, \& {Graham
  Smith}}]{1993MNRAS.265.1003L}
{Lyne}, A.~G., {Pritchard}, R.~S., \& {Graham Smith}, F. 1993, \mnras, 265,
  1003, \dodoi{10.1093/mnras/265.4.1003}

\bibitem[{{Manchester} {et~al.}(1982){Manchester}, {Tuohy}, \&
  {Damico}}]{1982ApJ...262L..31M}
{Manchester}, R.~N., {Tuohy}, I.~R., \& {Damico}, N. 1982, \apjl, 262, L31,
  \dodoi{10.1086/183906}

\bibitem[{{Martin-Carrillo} {et~al.}(2012){Martin-Carrillo}, {Kirsch},
  {Caballero}, {Freyberg}, {Ibarra}, {Kendziorra}, {Lammers}, {Mukerjee},
  {Sch{\"o}nherr}, {Stuhlinger}, {Saxton}, {Staubert}, {Suchy}, {Wellbrock},
  {Webb}, \& {Guainazzi}}]{2012A&A...545A.126M}
{Martin-Carrillo}, A., {Kirsch}, M.~G.~F., {Caballero}, I., {et~al.} 2012,
  \aap, 545, A126, \dodoi{10.1051/0004-6361/201116576}

\bibitem[{{Molkov} {et~al.}(2010){Molkov}, {Jourdain}, \&
  {Roques}}]{2010ApJ...708..403M}
{Molkov}, S., {Jourdain}, E., \& {Roques}, J.~P. 2010, \apj, 708, 403,
  \dodoi{10.1088/0004-637X/708/1/403}

\bibitem[{{Peng} {et~al.}(2026){Peng}, {Weng}, {Ge}, {Zhou},
  {G{\"u}gercino{\u{g}}lu}, {Ye}, {Tuo}, {Zhang}, {Zhang}, {Zheng}, {Zheng}, \&
  {Wang}}]{2026ApJ...999...85P}
{Peng}, H.-L., {Weng}, S.-S., {Ge}, M.-Y., {et~al.} 2026, \apj, 999, 85,
  \dodoi{10.3847/1538-4357/ae4007}

\bibitem[{{Perera} {et~al.}(2019){Perera}, {DeCesar}, {Demorest}, {Kerr},
  {Lentati}, {Nice}, {Os{\l}owski}, {Ransom}, {Keith}, {Arzoumanian}, {Bailes},
  {Baker}, {Bassa}, {Bhat}, {Brazier}, {Burgay}, {Burke-Spolaor}, {Caballero},
  {Champion}, {Chatterjee}, {Chen}, {Cognard}, {Cordes}, {Crowter}, {Dai},
  {Desvignes}, {Dolch}, {Ferdman}, {Ferrara}, {Fonseca}, {Goldstein},
  {Graikou}, {Guillemot}, {Hazboun}, {Hobbs}, {Hu}, {Islo}, {Janssen},
  {Karuppusamy}, {Kramer}, {Lam}, {Lee}, {Liu}, {Luo}, {Lyne}, {Manchester},
  {McKee}, {McLaughlin}, {Mingarelli}, {Parthasarathy}, {Pennucci}, {Perrodin},
  {Possenti}, {Reardon}, {Russell}, {Sanidas}, {Sesana}, {Shaifullah},
  {Shannon}, {Siemens}, {Simon}, {Spiewak}, {Stairs}, {Stappers}, {Swiggum},
  {Taylor}, {Theureau}, {Tiburzi}, {Vallisneri}, {Vecchio}, {Wang}, {Zhang},
  {Zhang}, {Zhu}, \& {Zhu}}]{2019MNRAS.490.4666P}
{Perera}, B.~B.~P., {DeCesar}, M.~E., {Demorest}, P.~B., {et~al.} 2019, \mnras,
  490, 4666, \dodoi{10.1093/mnras/stz2857}

\bibitem[{{Qi} {et~al.}(2025){Qi}, {Zheng}, {Zhang}, {Ge}, {Li}, {Zhang}, {Lu},
  {Peng}, {Zhang}, {Feng}, {Zhang}, {Xu}, {Li}, {Song}, {Zhang}, {Tao}, \&
  {Ye}}]{2025ApJ...981...99Q}
{Qi}, L., {Zheng}, S., {Zhang}, J., {et~al.} 2025, \apj, 981, 99,
  \dodoi{10.3847/1538-4357/adb42f}

\bibitem[{{Ray} {et~al.}(2019){Ray}, {Guillot}, {Ransom}, {Kerr}, {Bogdanov},
  {Harding}, {Wolff}, {Malacaria}, {Gendreau}, {Arzoumanian}, {Markwardt},
  {Soong}, \& {Doty}}]{2019ApJ...878L..22R}
{Ray}, P.~S., {Guillot}, S., {Ransom}, S.~M., {et~al.} 2019, \apjl, 878, L22,
  \dodoi{10.3847/2041-8213/ab2539}

\bibitem[{{Seward} \& {Harnden}(1982)}]{1982ApJ...256L..45S}
{Seward}, F.~D., \& {Harnden}, Jr., F.~R. 1982, \apjl, 256, L45,
  \dodoi{10.1086/183793}

\bibitem[{{Sheikh} {et~al.}(2006){Sheikh}, {Pines}, {Ray}, {Wood},
  {Lovellette}, \& {Wolff}}]{2006JGCD...29...49S}
{Sheikh}, S.~I., {Pines}, D.~J., {Ray}, P.~S., {et~al.} 2006, Journal of
  Guidance Control Dynamics, 29, 49, \dodoi{10.2514/1.13331}

\bibitem[{{Tuo} {et~al.}(2022){Tuo}, {Li}, {Ge}, {Nie}, {Song}, {Xu}, {Zheng},
  {Lu}, {Zhang}, {Liu}, {Cao}, {Chen}, {Qu}, {Zhang}, {Zhao}, {Xiao}, {Wu},
  {Wen}, {Jiang}, {Meng}, {Cui}, {Li}, {Zhang}, {Li}, {Yang}, {Tan}, \&
  {Li}}]{2022ApJS..259...14T}
{Tuo}, Y., {Li}, X., {Ge}, M., {et~al.} 2022, \apjs, 259, 14,
  \dodoi{10.3847/1538-4365/ac4250}

\bibitem[{{Yang} {et~al.}(2023){Yang}, {Wang}, {Han}, {Wang}, {Cui}, {Zhu},
  {Cong}, {Ma}, {Zhao}, {Hou}, {Yang}, {Chen}, {Lu}, {Lv}, {Sun}, {Zhang},
  {Yu}, {Wang}, {Liu}, {Zhang}, {Bi}, {Lu}, {Friedrich}, {Eder}, {Hartmann},
  {Burwitz}, {Keereman}, {Santovincenzo}, {Vernani}, {Bianucci}, {Valsecch},
  {Sheng}, {Yan}, {Qiang}, {Wang}, {Wang}, {Wang}, {Ding}, {Wang}, {Cheng}, \&
  {Chen}}]{2023ExA....55..625Y}
{Yang}, Y., {Wang}, Y., {Han}, D., {et~al.} 2023, Experimental Astronomy, 55,
  625, \dodoi{10.1007/s10686-022-09870-9}

\bibitem[{{Ye} {et~al.}(2025){Ye}, {Zhang}, {Ge}, {Qi}, {Zheng}, {Tuo}, {Han},
  {Peng}, {Xiao}, {Shui}, \& {Lu}}]{2025ApJ...993..108Y}
{Ye}, W.~T., {Zhang}, P.~F., {Ge}, M.~Y., {et~al.} 2025, \apj, 993, 108,
  \dodoi{10.3847/1538-4357/ae0303}

\bibitem[{{Yuan} {et~al.}(2022){Yuan}, {Zhang}, {Chen}, \&
  {Ling}}]{2022hxga.book...86Y}
{Yuan}, W., {Zhang}, C., {Chen}, Y., \& {Ling}, Z. 2022, in Handbook of X-ray
  and Gamma-ray Astrophysics, ed. C.~{Bambi} \& A.~{Sangangelo} (Springer), 86,
  \dodoi{10.1007/978-981-16-4544-0_151-1}

\bibitem[{{Zhang} {et~al.}(2025){Zhang}, {Chen}, {Jia}, {Zhao}, {Cui}, {Chen},
  {Wang}, {Wang}, {Wang}, {Li}, {Zhao}, {Guan}, {Han}, {Xu}, {Song}, {Feng},
  {Zhang}, \& {Yuan}}]{2025RAA....25k5019Z}
{Zhang}, J., {Chen}, Y., {Jia}, S., {et~al.} 2025, Research in Astronomy and
  Astrophysics, 25, 115019, \dodoi{10.1088/1674-4527/ae05f8}

\bibitem[{{Zhang}(2022)}]{2022MNRAS.511L..19Z}
{Zhang}, W. 2022, \mnras, 511, 19, \dodoi{10.1093/mnrasl/slab133}

\bibitem[{{Zhao} {et~al.}(2017){Zhao}, {Zhang}, {Yuan}, {Zhang}, {Willingale},
  \& {Ling}}]{2017ExA....43..267Z}
{Zhao}, D., {Zhang}, C., {Yuan}, W., {et~al.} 2017, Experimental Astronomy, 43,
  267, \dodoi{10.1007/s10686-017-9534-5}

\bibitem[{{Zhao} {et~al.}(2025{\natexlab{a}}){Zhao}, {Li}, {Wang}, {Zhang},
  {Jia}, {Guan}, {Zhao}, {Chen}, {Xu}, {Han}, {Song}, \&
  {Cui}}]{2025RDTM....9..215Z}
{Zhao}, H.-S., {Li}, C.-K., {Wang}, J., {et~al.} 2025{\natexlab{a}}, Radiation
  Detection Technology and Methods, 9, 215, \dodoi{10.1007/s41605-025-00526-8}

\bibitem[{{Zhao} {et~al.}(2025{\natexlab{b}}){Zhao}, {Cui}, {Wang}, {Zhang},
  {Zhao}, {Hou}, {Zhu}, {Wang}, {Xu}, {Luo}, {Han}, {Yang}, {Wang}, {Ma},
  {Yang}, {Huo}, {Li}, {Zhang}, {Geng}, \& {Chen}}]{2025RAA....25a5002Z}
{Zhao}, X., {Cui}, W., {Wang}, H., {et~al.} 2025{\natexlab{b}}, Research in
  Astronomy and Astrophysics, 25, 015002, \dodoi{10.1088/1674-4527/ad981b}

\bibitem[{{Zheng} {et~al.}(2024){Zheng}, {Han}, {Xu}, {Lee}, {Yuan}, {Wang},
  {Ge}, {Zhang}, {Li}, {Yin}, {Ma}, {Chen}, \& {Zhang}}]{2024Univ...10..174Z}
{Zheng}, S., {Han}, D., {Xu}, H., {et~al.} 2024, Universe, 10, 174,
  \dodoi{10.3390/universe10040174}

\end{thebibliography}
\bibliographystyle{aasjournal}


\end{CJK*}
\end{document}